\documentclass[letterpaper,10pt]{article}
\usepackage{osameet2}

\usepackage{amsmath,amssymb}
\usepackage{amsfonts}%

\usepackage{floatrow}
\newfloatcommand{capbtabbox}{table}[][\FBwidth]
\usepackage{blindtext}

\usepackage{lipsum}

\newlength\Colsep
\usepackage{multirow}
\usepackage[pdftex,colorlinks=true,bookmarks=false,citecolor=blue,urlcolor=blue]{hyperref} 

\begin{document}

\title{Machine Learning for QoT Estimation of Unseen Optical Network States}

\author{Tania Panayiotou$^{(1)*}$, Giannis Savva$^{(1)}$, Behnam Shariati$^{(2)}$,  Ioannis Tomkos$^{(3)}$, Georgios Ellinas$^{(1)}$}
\address{$^{(1)}$ KIOS CoE and the Dep. of Electrical and Computer Engineering,
University of Cyprus, Nicosia 1678, Cyprus \\
 $^{(2)}$ Universitat Polit\'{e}cnica de Catalunya (UPC), 08034 Barcelona, Spain \\ $^{(3)}$ Athens Information Technology, Marousi, 15125 Athens, Greece }
\email{$^*$ panayiotou.tania@ucy.ac.cy}

\begin{abstract}
We apply deep graph convolutional neural networks for Quality-of-Transmission estimation of unseen network states capturing, apart from other important impairments, the inter-core crosstalk that is prominent in optical networks operating with multicore fibers.
\end{abstract}


\section{Introduction}
Traditional network planning tools estimate Quality-of-Transmission (QoT) with static Q-factor models~\cite{Shariati18} which are functions of the physical layer impairments (PLIs). The calculation of these PLIs requires complex computations, time-consuming measurements, and expensive equipment that take considerable human effort and lack self-adaptivenes. Moreover, these models are usually based on worst-case budget values for accounting for the non-linear PLIs that are difficult to be accurately evaluated upfront~\cite{Panayiotou17}. This, however, results in inefficient utilization of the spectrum resources, especially when the worst-case PLIs are highly overestimated~\cite{Klinkowski18b}. To alleviate these problems, machine learning (ML) techniques have been recently applied for inferring, from the QoT data of previously established connections, QoT models that are not a function of the PLIs and are robust to network changes~\cite{Panayiotou17, Morais18, Proietti18}. Among the ML methods applied (K-nearest neighbors, logistic regression, support vector machines, and neural networks (NNs)), NNs have shown to present better generalization and higher accuracy~\cite{Morais18}. NNs' accuracy is also experimentally demonstrated in~\cite{Proietti18}, while a software defined networking (SDN)-based implementation scheme incorporating the data-driven QoT estimation framework is presented in~\cite{Chen18}. 

Related work~\cite{Panayiotou17, Morais18, Proietti18}, to the best of our knowledge, apply ML aiming at finding QoT estimation models that are a function only of the new (unestablished) lightpath. Consequenlty, these models cannot capture the crosstalk (XT) effect between the new and the established lightpahts. To achieve this, the models must be a function not only of the new lightpath, but also of the interractions between the new and the already established lightpaths; a signficant consideration, especially for the spectrally-spatially flexible optical networks (SS-FONs) with multicore fibers (MCFs) where the inter-core XT may severely affect the QoT of both the new and the already established lightpaths. In this work, we advance the state-of-the art by applying a deep graph convolutional NN (DGCNN)~\cite{Zhang18} capable of finding a QoT model that is a function of a network state capturing: (1) the lightpaths' dependencies and (2) the lightpaths' specific features. The aim is for the inferred QoT model to accurately classify any unseen network state (not used during learning) to one of two classes; the feasible QoT or the infeasible QoT class. Note that the applicability of NNs on graphs (i.e., temporal network states), in computational inexpensive ways, has only recently been made possible~\cite{Zhang18}.
\section{Deep Graph Convolutional Neural Networks for QoT Estimation}
The QoT estimation problem is formulated according to a DGCNN for a dynamic SS-FON with MCFs: \\ 
$\blacktriangleright$ An SS-FON with MCFs is given by $\Gamma(V,E,K,B,n)$, where $V$ is the set of network nodes, $E$ is the set of fiber links, $K=\{\kappa\}_{\kappa=1}^{|K|}$ is the core identification set (all fiber links have $|K|$ cores), $B=\{b\}_{b=1}^{|B|}$ is the spectrum slot identification set (each core has $|B|$ slots), and $n$ is the number of all possible combinations of source-destination pairs (connections) in $\Gamma$ (each connection is indexed with a different number $i$, where $i=1,..,n$).\\
$\blacktriangleright$ A DGCNN, given a collection of graphs in the form $(G,y)$, where $G$ is a graph and $y$ is its class, learns a function that can be used for classification on unseen graphs~\cite{Zhang18}. A graph $G$ is described by a symmetric adjacency matrix ${\bf A}$ and a $c$-dimensional feature vector ${\bf X} \in \mathbb{R}^{n \times c}$, where $n$ is the number of nodes in $G$ and ${\bf X}$ denotes the graph's node information matrix, with each row representing a vertex. 
\subsection{Problem Formulation} 
We define a dataset $D=({\bf G},{\bf y})=\{G^{s}({\bf A}^s,{\bf X}^s), y^s\}_{s=1}^N$ where, $G^s$ is the state that the network may transition to if the $s^{th}$ connection request is established. ${\bf A}^s$ is the adjacency matrix of $G^s$, and ${\bf X}^s$ is the feature vector of $G^s$. The aim is to infer from $D$, a QoT model, $f(G^{s})=y^s \in [0,1]$, that accurately estimates the feasibility of the $G^{s}$ network state. $G^{s}$ is feasible if the QoT of all the lightpaths at $G^{s}$ is above a predetermined QoT threshold; in that case $y^s=1$, otherwise $y^s=0$. For achieving our problem objective, ${\bf A}^s$ and ${\bf X}^s$, are defined according to what we can know about $G^{s}$, upon $s$: \\ 
$\blacktriangleright$ ${\bf A}^s:$ a symmetric matrix of size $n \times n$, where $A^{s}_{ij}=1$ if the ligthpaths of the connections indexed with $i$ and $j$ will share at least one common link at $G^{s}$, otherwise $A^{s}_{ij}=0$. In essence, ${\bf A}^s$, captures the information related to which lightpaths will be susceptible to XT at $G^{s}$, without having to describe the full routes of the lightpaths. This is very important as by doing so the QoT estimation process is independent from the exact routing procedure followed (i.e., the problem is not formulated according to a set of routes that are a priori restricted for each connection). This in turn, means that the network performance cannot be affected by the problem formulation.     
\\ $\blacktriangleright$ ${\bf X}^s:$ a matrix of size $n \times c$ and describes the features of each lightpath at $G^{s}$. In particular, row $i$ corresponds to the feature vector ${\bf  x^s_i}=[x^s_{i1},..,x^s_{ic}]$ and describes the lightpath of the $i^{th}$ connection at $G^{s}$. Thus, we set $\{x^s_{ij}=0\}_{j=1}^{c}$ if the connection $i$ will not be established at $G^{s}$ (it is not established in the previous network state, $G^{s'}$, and does not correspond to the $s^{th}$ arrival), otherwise:
%
$(a)$ $x^s_{i1} \in \mathbb{R}$ is the entire length (in km) of $i$ at $G^{s}$,
$(b)$ $x^s_{i2} \in \mathbb{R}$ is the maximum link length (in km) of $i$ at $G^{s}$, $(c)$ $x^s_{i3} \in B$ is the central frequency (slot number) allocated to $i$ at $G^{s}$, 
$(d)$ $x^s_{i4} \in \mathbb{R}$ is the number of slots allocated to $i$ at $G^{s}$, 
$(e)$ $x^s_{i5} \in K$ is the core number allocated to $i$ at $G^{s}$ (the same core is allocated along the entire $i$ - a core allocation scheme that allows core switching is planned for future work), 
$(f)$ $x^s_{i6} \in [1,2,3,4]$, $x^s_{i6}=1,2,3,4$ if BPSK, QPSK, 8-QAM or 16-QAM modulation format is used for $i$ at $G^{s}$, 
$(g)$ $x^s_{i7} \in \mathbb{R}$ is the number of EDFAs along $i$ at $G^{s}$, 
$(h)$ $x^s_{i8} \in \mathbb{R}$ is the number of links along $i$ at $G^{s}$, 
$(i)$ $x^s_{i9} \in \mathbb{R}$ is a QoT indication of $i$ at $G^{s'}$. In particular, $x^s_{i9}=0$ if $i$ corresponds to the $s^{th}$ connection request, otherwise $x^s_{i9}= BER^s_i$, where $BER^s_i$ is the bit-error-rate (BER) of $i$ at $G^{s'}$ (before the $s^{th}$ connection is established). These values are used because the BER at the receiver nodes is known only for the already established connections (through OPM~\cite{Proietti18, Chen18}).
\subsection{Data-Driven QoT Framework}
\begin{figure}[h!]
\begin{center}
\includegraphics[scale=0.42]{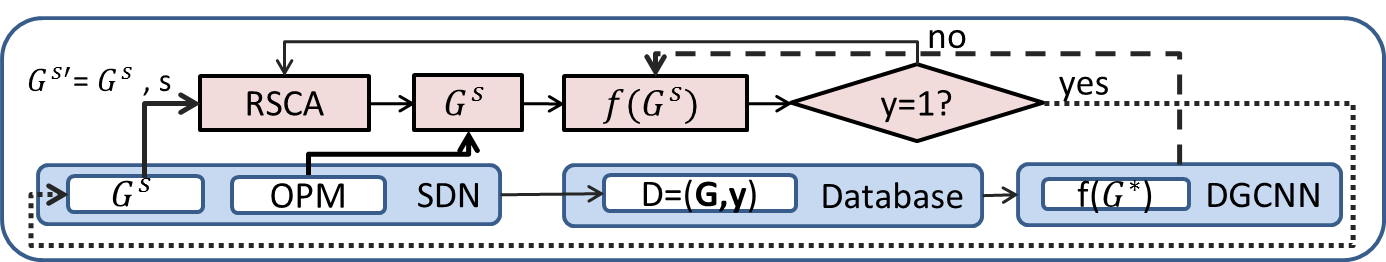}
\vspace{-0.17in}
\caption{Data-driven QoT framework}
\label{res}
\end{center}
\end{figure}

An abstracted framework incorporating the ML application is illustrated in Fig.~\ref{res}, where an optical network is centrally controlled by an SDN-based  controller~\cite{Proietti18} that dynamically monitors
and configures the data plane. A database platform collects and stores real-time network state information and optical network monitoring information through OPM~\cite{Proietti18, Chen18}. Specifically, it collects and stores information capable of fully describing any network state, $G^s$, and its ground truth $y^s$. $G^s$ is computed upon the arrival of each connection request, $s$, by a routing, spectrum and core allocation (RSCA) procedure, given as input the previous network state, $G^{s'}$. If the network transitions to $G^{s}$, OPM is used to find the real BER values (ground truth). It is essential that probing lightpaths or alien wavelengths~\cite{Proietti18, Chen18} are used for enriching the dataset information (especially for the infeasible network states). 
\vspace{-0.09in}
\vspace{-0.18in}

The ML application runs on top of the database. It retrieves the $D=({\bf G}, {\bf y})$ dataset used for finding the QoT model, $f(G^{*})$, by training the DGCNN. After model convergence, $f(G^{*})$ classifies any unseen network state, $G^{*}$. In particular, upon any $s$, if $f(G^{s})=1$, then the network transitions to $G^{s}$. Otherwise, depending on the RSCA procedure followed, $s$ may be either blocked, or an alternative lightpath may be tested for $s$. Note that the QoT model can be offline retrained according to the most recent dataset, for capturing the network changes (i.e., network degradation). Retraining can be performed either periodically or after the controller observes a number of misclassifications. For avoiding the misclassification side-effects, a safety margin could be used in conjunction with the QoT model.
\section{Dataset Generation, Model Training and Accuracy}
\begin{figure}[H]
\begin{floatrow}
\capbtabbox[3.8in][3.5cm]{%
\footnotesize
\begin{tabular}{|c|c|c|c|c|c|c|}
\hline
\multirow{2}{*}{Case} & \multicolumn{3}{c|}{Number of patterns} & \multirow{2}{*}{ACC ($\%$)} & \multirow{2}{*}{AUC} & \multirow{2}{*}{Training time (mins)} \\ \cline{2-4}
                      & $\boldsymbol{D}$        & $\boldsymbol{D}_{trn}$        & $\boldsymbol{D}_{ts}$      &                           &                      &                                      \\ \hline
1                     & 3000           & 2700       & 300       & 97                        & 0.98                 & 86                                   \\ \hline
2                     & 1800           & 1620       & 180       & 93                        & 0.96                 & 59                                   \\ \hline
3                     & 1500           & 1350       & 150       & 92                        & 0.95                 & 36                                   \\ \hline
\end{tabular}
}{%

  \caption{Datasets Information, Model Accuracy, and Training Time}%
  \label{tab1}
  \vspace{-0.1in}
}
\ffigbox[2.4in]{
\includegraphics[scale=0.24]{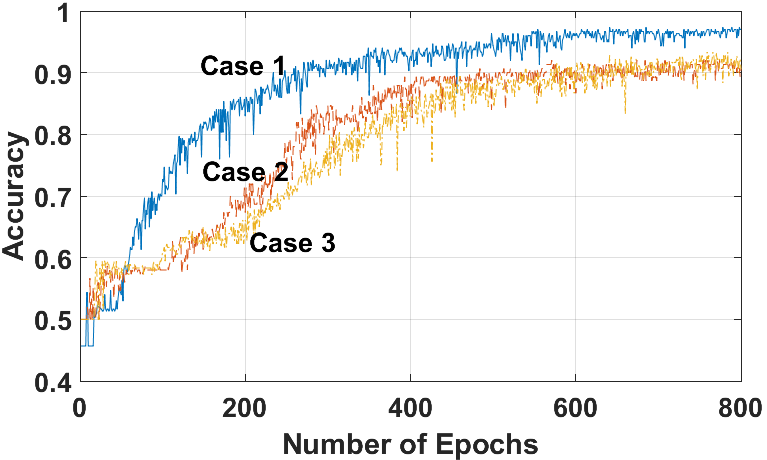}%
}{\caption{Accuracy vs. the number of epochs}%
\label{figres}
  \vspace{-0.1in}
}
\end{floatrow}
\end{figure}

The national Telefonica network topology (30 nodes, 56 undirected links) is assumed, with each fiber link having a capacity of 160 spectrum slots and each fiber being a single-mode 7 core fiber. In total 20000 requests were generated in a dynamic optical network according to the Poisson process with exponentially distributed holding times for a netork load of 400 Erlangs. For each request, $s$, we extracted its ($G^s$,$y^s$) pattern. Each $s$ was associated with a specific source-destination pair by randomly sampling between the set of possible source-destination pairs (sampled from the range of numbers $[1, n]$). Each $G^{s}$ was found accroding to a conventional RSCA heuristic~\cite{Savva18}, and according to the previous network state, $G^{s'}$ (Fig.~\ref{res}). Ground truth $y^s$ was generated by the Q-factor model~\cite{Shariati18}. Q-threshold was set to a corresponding BER of $10^{-3}$~\cite{Shariati18}. As shown in Table~\ref{tab1}, 3 datasets $\boldsymbol {D}$ were created, that differ in their size.  
 
For training the QoT model we used a similar DGCNN configuration as the one described in~\cite{Zhang18}. For each $\boldsymbol {D}$ case (Table~\ref{tab1}), we performed 10-fold cross validation. Each fold is created by sampling patterns from their respective $\boldsymbol{D}$. Information regarding the training ($\boldsymbol{D}_{trn}$) and test ($\boldsymbol{D}_{tst}$) datasets is shown in Table~\ref{tab1}. Note that for each case examined, $\boldsymbol{D}_{trn}$ and $\boldsymbol{D}_{tst}$ are completely different sets. Also, the number of patterns for which $y=1$ and $y=0$ is balanced in all sets (i.e., 150 patterns with $y=1$, 150 with $y=0$ in $\boldsymbol{D}_{tst}$). The DGCNN was trained according to the ADAM algorithm~\cite{Zhang18}. The accuracy (ACC), within each $\boldsymbol{D}_{tst}$ case, versus the number of epochs is shown in Fig.~\ref{figres} clearly showing that the DGCNN convergences to a QoT model of an ACC ranging between $92\%-97\%$. Table~\ref{tab1} reports the training time, the average ACC and the average area under the curve (AUC) obtained within each (unseen) $\boldsymbol{D}_{tst}$, after model convergence (AUC indicates whether the ACC achieved within each class separately is balanced and it is desirable that the AUC is close to $1$). Note that a PC with a GPU1050-TI and 24 GB RAM was used. After model training, a graph is classified in milliseconds. According to Table~\ref{tab1}, an accurate QoT model is obtained in practical time for each case examined.
\vspace{-0.2in}
\vspace{-0.15in}

Overall the results indicate that the proposed ML method is a promising candidate for QoT estimation purposes, alleviating the static Q-factor problems. Importantly, it explicitly considers, during inference, the XT effect and it does so in a way that does not affect other network functionalities (i.e., routing decisions).
To the best of our knowledge there is no other supervised ML technique (other than applying graph NNs) capable of achieving the same objective. Any other NN requires describing a $G^{s}$ state through a single $\bf x^{s}$ vector; meaning that specific elements of $\bf x^{s}$ must be devoted for describing a priori computed routes (restricting routing decisions). Experimental demonstration of the proposed approach along with examining any related implementation issues constitute interesting future research.  
\section*{Acknowledgement}This work has been supported by the European Union’s Horizon 2020 research and innovation programme under grant agreement No 739551 (KIOS CoE- Teaming) and from of the Republic of Cyprus through the Directorate General for European Programmes, Coordination and Development.

\end{document}